\documentclass[fleqn,usenatbib]{mnras}

\usepackage{newtxtext,newtxmath}
\usepackage[T1]{fontenc}

\DeclareRobustCommand{\VAN}[3]{#2}
\let\VANthebibliography\thebibliography
\def\thebibliography{\DeclareRobustCommand{\VAN}[3]{##3}\VANthebibliography}

\usepackage{graphicx}	% Including figure files
\usepackage{amsmath}	% Advanced maths commands

\title{Propagation and Energy Dissipation of Shock Waves in the Solar Chromosphere }

\author[Chaurasiya et al.]{
	Ravi Chaurasiya,$^{1,2}$\thanks{E-mail: ravi@prl.res.in}
	Ankala Raja Bayanna,$^{1}$
	Robertus Erdélyi$^{3,4,5}$
	\\
	$^{1}$Udaipur Solar Observatory, Physical Research Laboratory, Udaipur-313001, India \\
	$^{2}$ Indian Institute of Technology, Gandhinagar, Gujarat-382355, India\\
	$^{3}$Solar Physics and Space Plasma Research Centre (SP2RC), School of Mathematical and Physical Sciences, University of Sheffield, Sheffield S3 7RH, UK\\
	$^{4}$Department of Astronomy, Eötvös Loránd University, Budapest, Pázmány P. sétány 1/A, H-1117, Hungary\\
	$^{5}$Gyula Bay Zoltan Solar Observatory (GSO), Hungarian Solar Physics Foundation (HSPF) Petőfi tér 3., Gyula H-5700, Hungary}

\date{Accepted XXX. Received YYY; in original form ZZZ}

\pubyear{\the\year{}}

\begin{document}
\label{firstpage}
\pagerange{\pageref{firstpage}--\pageref{lastpage}}
\maketitle

\begin{abstract}

The solar atmosphere is permeated by various types of waves that originate from subsurface convection. As these waves propagate upward, they encounter they encounter a steep decrease in the density of the medium, leading to their steepening into shock waves. These shock waves typically exhibit a characteristic sawtooth pattern in wavelength-time ($\lambda$-t) plots of various chromospheric spectral lines, viz., H$\alpha$, Ca II 8542 \AA~ to name a few. In this study, we investigate the propagation of shock waves in the lower solar atmosphere using coordinated observations from the Swedish 1-meter Solar Telescope (SST), the Interface Region Imaging Spectrograph (IRIS), and the Solar Dynamics Observatory (SDO). Our analysis reveals that after forming in the chromosphere, these shock waves travel upward through the solar atmosphere, with their signatures detectable not only in the transition region but also in low coronal passbands. These shock waves dissipate their energy into the chromosphere as they propagate. In certain cases, the energy deposited by these waves is comparable to the radiative losses of the chromosphere, highlighting their potential role in chromospheric heating. Our findings reported here provide crucial insights into wave dynamics in the lower solar atmosphere and their contribution to the energy transport process in the chromosphere.

\end{abstract}


%%%%%%%%%%%%%%%%%%%%%%%%%%%%%%%%%%%%%%%%%%%%%%%%%%

%%%%%%%%%%%%%%%%% BODY OF PAPER %%%%%%%%%%%%%%%%%%

\section{Introduction} \label{Sec1}

The solar atmosphere comprises the photosphere, chromosphere, transition region, and corona. Acting as a bridge between the cooler photosphere and the hotter corona, the chromosphere is slightly hotter than the photosphere but much denser than the corona. Consequently, it requires more energy to compensate for radiative losses, ranging from \(10^6\)–\(10^7\)~erg~cm\(^{-2}\)~s\(^{-1}\), compared to the corona’s \(10^4\)–\(10^6\)~erg~cm\(^{-2}\)~s\(^{-1}\) \citep{1977ARA&A..15..363W,1989ApJ...336.1089A}.

The solar atmosphere exhibits a highly dynamic nature, hosting various jet-like structures such as spicules (\cite{2004Natur.430..536D,2014ApJ...792L..15P,2019ARA&A..57..189C,2022NatPh..18..595D,2024ApJ...970..179C,2024ApJ...973...49K}), alongside a wide range of wave phenomena (\cite{2009SSRv..149..355Z,2013SSRv..175....1M,2015SSRv..190..103J,2023LRSP...20....1J,2025MNRAS.537.2243C,2025ApJ...989..129S}).
Spicules are thin, hair-like structures readily observed at the solar limb in chromospheric spectral lines. The on-disk counterparts of these spicules are referred as fibrils/mottles (\cite{1995ApJ...450..411S,2006ApJ...647L..73H,2013ApJ...779...82K}). These structures undergo upward and downward motions within the solar atmosphere and can be heated to transition region and coronal temperatures (\cite{2011Sci...331...55D,2014ApJ...792L..15P,2015ApJ...799L...3R,2016ApJ...820..124H,2019Sci...366..890S,2024ApJ...970..179C}). Studies by \cite{2015ApJ...815L..16S} and \cite{2017ApJ...845L..18D} have also linked the spicules with the propagating disturbances seen in the space time maps of the coronal passbands.

Meanwhile, waves in the solar atmosphere predominantly originate in the photosphere or even deeper layers due to convective motions. These waves are ubiquitous  and present an important aspect to study in the solar atmosphere as these waves have the potential to contribute to chromospheric and coronal heating problem (\cite{2009Sci...323.1582J,2019NatCo..10.3504L}). As these waves propagate upward, they experience a rapid decline in atmospheric density, leading to a substantial increase in their amplitudes. To conserve energy, the waves eventually steepen and evolve into shock waves. These shock waves are seen as sawtooth wave pattern in wavelength-time ($\lambda$-t) plots of chromospheric spectral lines, such as H$\alpha$ and Ca II lines (\cite{1994chdy.conf...47C,1997ApJ...481..500C,2003A&A...403..277R,2023LRSP...20....1J}).

\cite{1997ApJ...481..500C} have suggested that the bright grain observed in Ca II H$_{2V}$ are a consequence of upward-propagating acoustic shock waves and appears as a sawtooth wave pattern in the $\lambda$-t plot of Ca II H line. Their results were based on the comparsion of the observation with the simulation of bright grains in Ca II H line using  one-dimensional hydrodynamic models with radiation in non-local thermodynamic equilibrium (non-LTE) assuming complete redistribution (CRD). Based on the idea that the presence of a sawtooth pattern serves as an observational signature of chromospheric shock waves, \cite{2003A&A...403..277R} demonstrated that such shock waves are predominantly located in the umbra of sunspots.

Decreasing density would amplify their amplitude, as shock waves propagate upward through the solar atmosphere, making them detectable in the transition region. However, \citet{1997ApJ...490L.195J,1997ESASP.404..685S}, using SUMER \citep{1995SoPh..162..189W} onboard SOHO \citep{domingo1995soho}, found no transition region counterparts for bright grains, suggesting that acoustic shock waves do not significantly heat the transition region or beyond. Later, \citet{2009A&A...494..269V} proposed that horizontal magnetic fields forming a canopy may block acoustic wave propagation. With coordinated SST \citep{2003SPIE.4853..341S} and IRIS \citep{2014SoPh..289.2733D} observations, \citet{2015ApJ...803...44M} found that some bright grains show Si IV 1403~Å signatures, indicating that certain acoustic shocks can reach the transition region.

\citet{2014ApJ...786..137T} and \citet{2021ApJ...906..121K}, using IRIS observations, showed that shock waves in the umbral transition region often exhibit sawtooth patterns in $\lambda$-t diagrams. \citet{2019ARA&A..57..189C} further suggested that the propagation of acoustic shocks into the upper atmosphere depends strongly on the magnetic field topology, determining whether they are refracted, reflected, or transmitted.

Despite numerous independent detections of shocks in the chromosphere and transition region, coordinated observations tracing their propagation from the lower chromosphere to the corona remain rare. Thus, it remains unclear whether all chromospheric shocks reach higher layers or some reflect back. In this study, we use coordinated observations from SST, IRIS, and (SDO, \cite{2012SoPh..275....3P})/Atmospheric Imaging Assembly (AIA, \cite{2012SoPh..275...17L}) and SDO/Helioseismic Magnetic Imager (HMI,\space \cite{2012SoPh..275..207S}) to investigate shock wave propagation from the low chromosphere through the transition region into coronal heights. This approach helps explore the coupling between atmospheric layers, the origin of spicules \citep{2015ApJ...815L..16S,2017ApJ...845L..18D}, and the link between shocks and coronal disturbances, which serves as a key motivation for this work. We also estimate the energy deposition by shocks in the chromosphere and compare it with radiative losses.

The rest of the paper is structured as follows: Section \ref{Sec2} details the observations used for this study. Section \ref{Sec3} presents the results, which include the detection of shock waves at multiple heights in the solar atmosphere at the same spatial location. These results are discussed and concluded in Section \ref{Sec4}.

\section{Observations} \label{Sec2}

\begin{figure*}
	\centering
	\includegraphics[width=170mm]{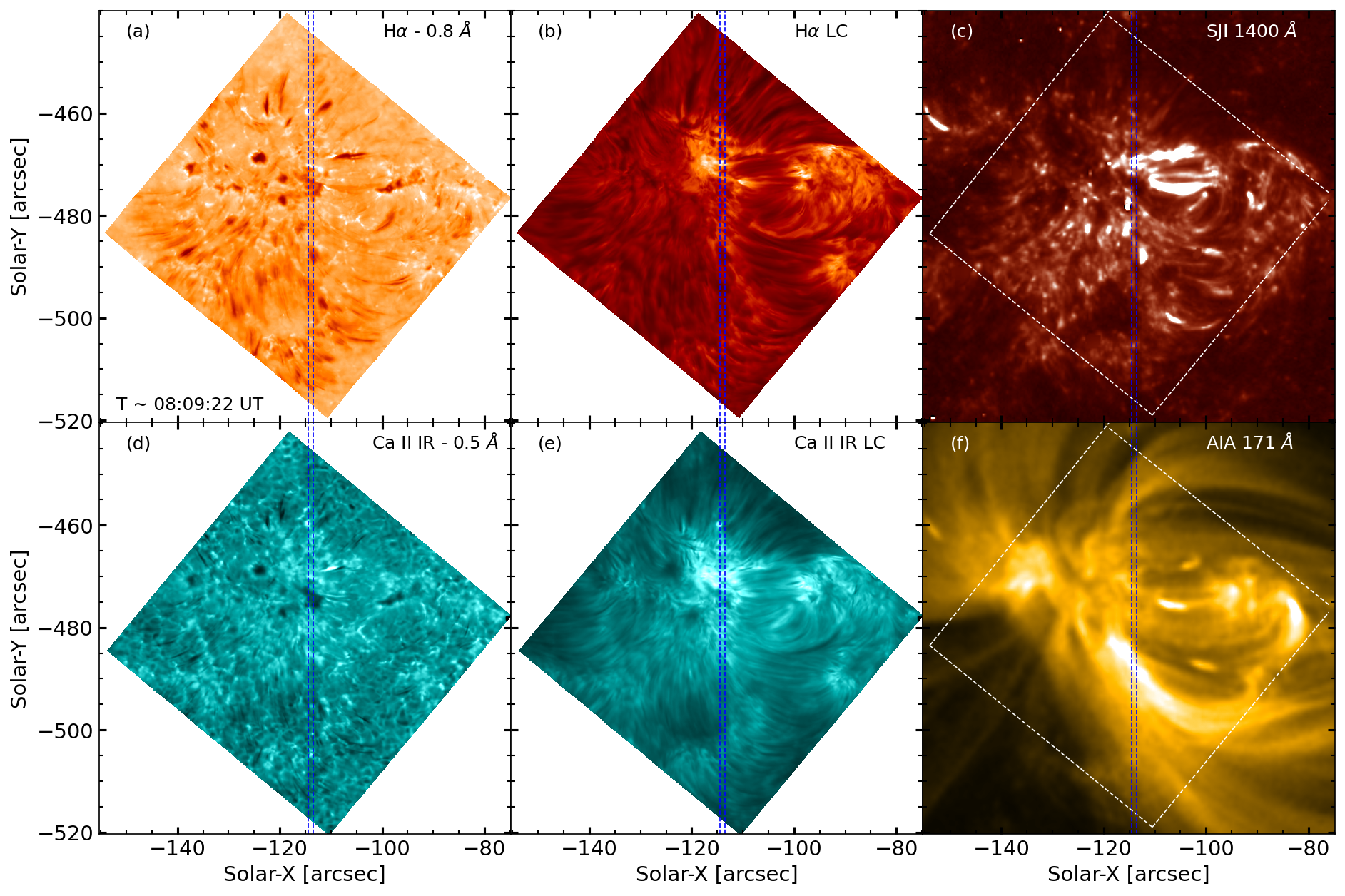}
	\caption{Panels (a) and (b) represent the blue wing and line center filtergrams of H$\alpha$, while panels (c) and (d) correspond to the blue wing and line center filtergrams of Ca~II~IR. Panels (e) and (f) illustrate the near-simultaneous appearance of the coordinated transition region and corona.
	 }
	\label{Fig1}
\end{figure*}

We analyzed a coordinated dataset of NOAA Active Region (AR) 12775, observed on 11 October 2020 from 08:04 UT to 08:35 UT using the SST, IRIS, SDO. The heliocentric coordinates of the target AR were ($-114\arcsec$, $-479\arcsec$), with a corresponding observing angle of $\mu = \cos\theta = 0.86$, where $\theta$ represents the heliocentric angle. This active region consisted of multiple pores, network regions with moderate magnetic field configuration, illustrated in Figure \ref{Fig1}. A part of this dataset has also been previously analysed by \cite{2024ApJ...970..179C} to study the multithermal nature of the spicules in the solar atmosphere.

\subsection{Swedish Solar Telescope (SST)} \label{sec2.1}

The SST dataset comprises imaging spectroscopic observations of the H$\alpha$, H$\beta$, and Ca II 8542 \AA~ spectral lines. The H$\alpha$ and Ca II 8542 \AA~ observations were obtained using the CRisp Imaging SpectroPolarimeter (CRISP, \cite{scharmer2008crisp}), while the H$\beta$ observations were recorded with the CHROmospheric Imaging Spectrometer (CHROMIS, \cite{scharmer2017solarnet}). In this study, we focus exclusively on the H$\alpha$ and Ca II 8542 \AA~ data. The H$\alpha$ observations span approximately 30 minutes, with a temporal cadence of 34.252 s. The field of view (FoV) of the dataset is around $60\arcsec \times 60\arcsec$, with a spatial sampling of $0\arcsec.0591$ per pixel. The spectral scans cover 17 positions in the H$\alpha$ line and 15 positions in the Ca II 8542 \AA~ line, with a spectral range of $\pm2$ Å and $\pm1.3$ Å, respectively. To ensure accurate spatial alignment, both datasets were coaligned by cross-correlating nearly simultaneous wing images dominated by photospheric features. The SST dataset analyzed in this work is publicly accessible through the SST archive\footnote{\url{https://dubshen.astro.su.se/sst_archive/search}}.

\subsection{Interface Region Imaging Spectrograph (IRIS)}

The coordinated IRIS\footnote{IRIS OBSID: 3630108417} Level 2 data (L12-2019-08-08 with \texttt{iris\_prep} version~2.67, run on 2022-09-01 at 11:57:18 and 12:55:11 UT), including both spectral data and Slit-Jaw Images (SJIs), were obtained from the SST–IRIS campaign encompassing the period of observations mentioned in Section \ref{sec2.1} above. The SJIs were recorded in three different wavelength channels: Si IV 1400 Å, Mg II k 2796 Å, and the Mg II h wing at 2832 Å, all with a uniform plate scale of 0.3327$''$. The average temporal cadence for these images was 21.88 s, 18.22 s, and 109.35 s, respectively. It should be noted that while the average temporal cadence of the \ion{Si}{iv}~1400~\AA\ observation is 21.88s, the cadence between successive images is not uniform. In fact, every sixth image in the \ion{Si}{iv}~1400~\AA\ sequence is taken at approximately double the interval of the preceding five images, which have a cadence of about 18.22s. To ensure a uniform cadence throughout the dataset, linear interpolation was applied to every sixth image, effectively resampling the entire \ion{Si}{iv}~1400~\AA\ time series to a constant cadence of 18.22s.

In addition to the SJIs, IRIS performed spectral observations using a large dense 4-step raster with a step size of 0.35$''$, resulting in a step cadence of 9.2 s and a full raster cadence of approximately 36 s. The spectroscopic data included emission from transition region and chromospheric lines: C II 1336 Å , Si IV 1394 Å and Si IV 1403 Å, and Mg II k 2796 Å. The spatial sampling along the slit was 0.3327$''$, while the spectral sampling was 50.91\,m\AA\ for the near-UV (Mg\,\textsc{ii}\,k) window, and 25.44\,m\AA\ (Si\,\textsc{iv}) and 25.96\,m\AA\ (C\,\textsc{ii}) for the far-UV windows. The wavelength calibration performed in the Level 2 standard IRIS pipeline utilizes neutral lines that are known to form in the photosphere or lower chromosphere and typically exhibit intrinsic velocities of less than 1~km\,s$^{-1}$
\citep{2014SoPh..289.2733D}. Specifically, the O I 1355.60 Å line was used for calibrating the far-UV spectral windows, while Ni I 2799.474 Å was employed for the near-UV calibration. 

To ensure accurate alignment between the SST and IRIS-SJI images, the SST data were first degraded to match the IRIS-SJI plate scale. Following this, a cross-correlation was performed between the nearly simultaneous IRIS 2832 Å SJI image and the photospheric wing image from SST H$\alpha$.

\subsection{Solar Dynamics Observatory (SDO)}

The registered coordinated cutouts were obtained using the SDO cutout sequence, which provides EUV data from SDO/AIA as well as continuum images and magnetograms from SDO/HMI. The AIA-UV channels (1600~\AA, 1700~\AA) and AIA-EUV channels (171~\AA, 193~\AA) had an approximately identical plate scale of 0$\arcsec$.6 pixel$^{-1}$, but their temporal cadences differed, with the EUV channels having a cadence of 12~s and the UV channels having a cadence of 24~s. The HMI data, including the line-of-sight (LOS) magnetograms and continuum images, had a plate scale of 0$\arcsec$.5 pixel$^{-1}$ and a temporal cadence of 45~s. To achieve precise alignment between the SDO images and the IRIS SJI observations, the individual AIA channels were first coaligned with each other. These coaligned AIA images were then bilinearly interpolated to match the IRIS SJI plate scale. Following this, a cross-correlation was performed between the nearly simultaneous AIA 1700~\AA{} and IRIS-SJI 2832~\AA{} images to establish the coordinated alignment of SDO/AIA with IRIS-SJIs observations. A similar procedure was applied to the HMI continuum images to get coalingned HMI continuum and magnetogram data.  

Figure~\ref{Fig1} presents an overview of the observations, displaying filtergrams of H$\alpha$ and Ca II 8542 \AA~ (both in the blue wing and at the line center), along with the coaligned IRIS SJI 1400~\AA{} and AIA 171~\AA{} images. The dashed vertical blue lines indicate the IRIS slit coverage across the field of view (FoV).

\section{Results}  \label{Sec3}

Building on previous studies and the established notion that the presence of sawtooth behavior in the $\lambda$-t plot serves as a key observational signature of shock waves in the solar atmosphere \citep{1997ApJ...481..500C,2003A&A...403..277R}, we investigate shock waves in a weak active region of the solar atmosphere. This study utilizes coordinated observations from the SST, IRIS and SDO. Specifically, we analyze the Stokes I parameters of the H$\alpha$ line, along with both Stokes I and Stokes V parameters of the Ca II 8542 \AA~ spectral line, to identify the locations of shock waves in the chromosphere. The results from this analysis are presented in Section \ref{Sec3.1}. Furthermore, we explore the transition region counterparts of those chromospheric shock waves which coincides with the slit positions of IRIS spectra, with results detailed in Section \ref{Sec3.2}. To investigate the impact of these shock waves beyond the chromosphere, we analyze light curves spanning the chromosphere, transition region, and corona to track the propagation of chromospheric shock waves into higher atmospheric layers. This analysis is discussed in Section \ref{Sec3.3}. Lastly, we estimate the shock wave flux deposited into the chromosphere using the information obtained from the STockholm inversion Code (STiC; \cite{2016ApJ...830L..30D,2019A&A...623A..74D}) to assess energy balance and wave dissipation effects, as detailed in Section \ref{Sec3.4}.

\subsection{Shock waves in the chromosphere} \label{Sec3.1}

\begin{figure*}
	\centering
	\includegraphics[width=90mm]{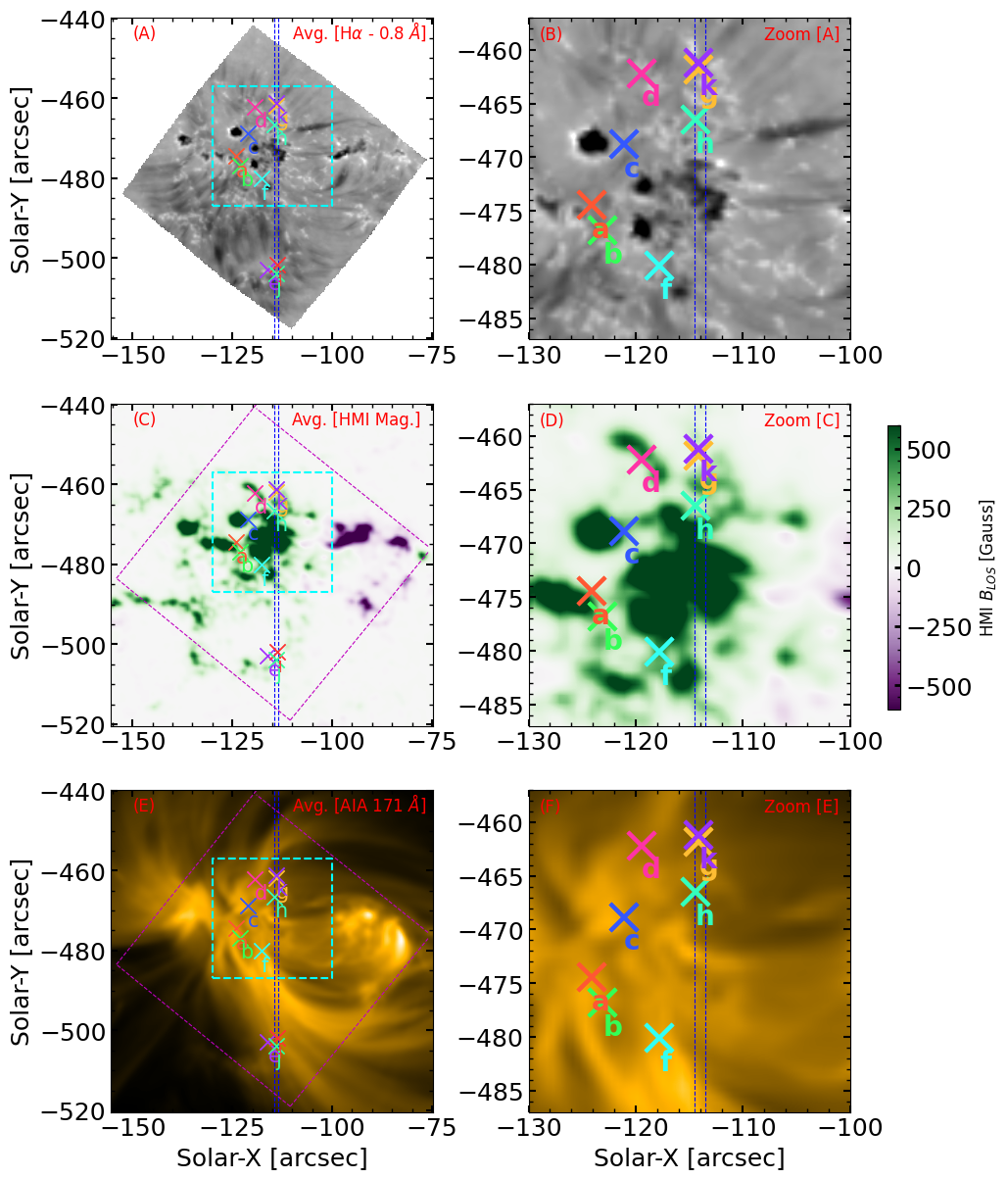}
	\includegraphics[width=81mm]{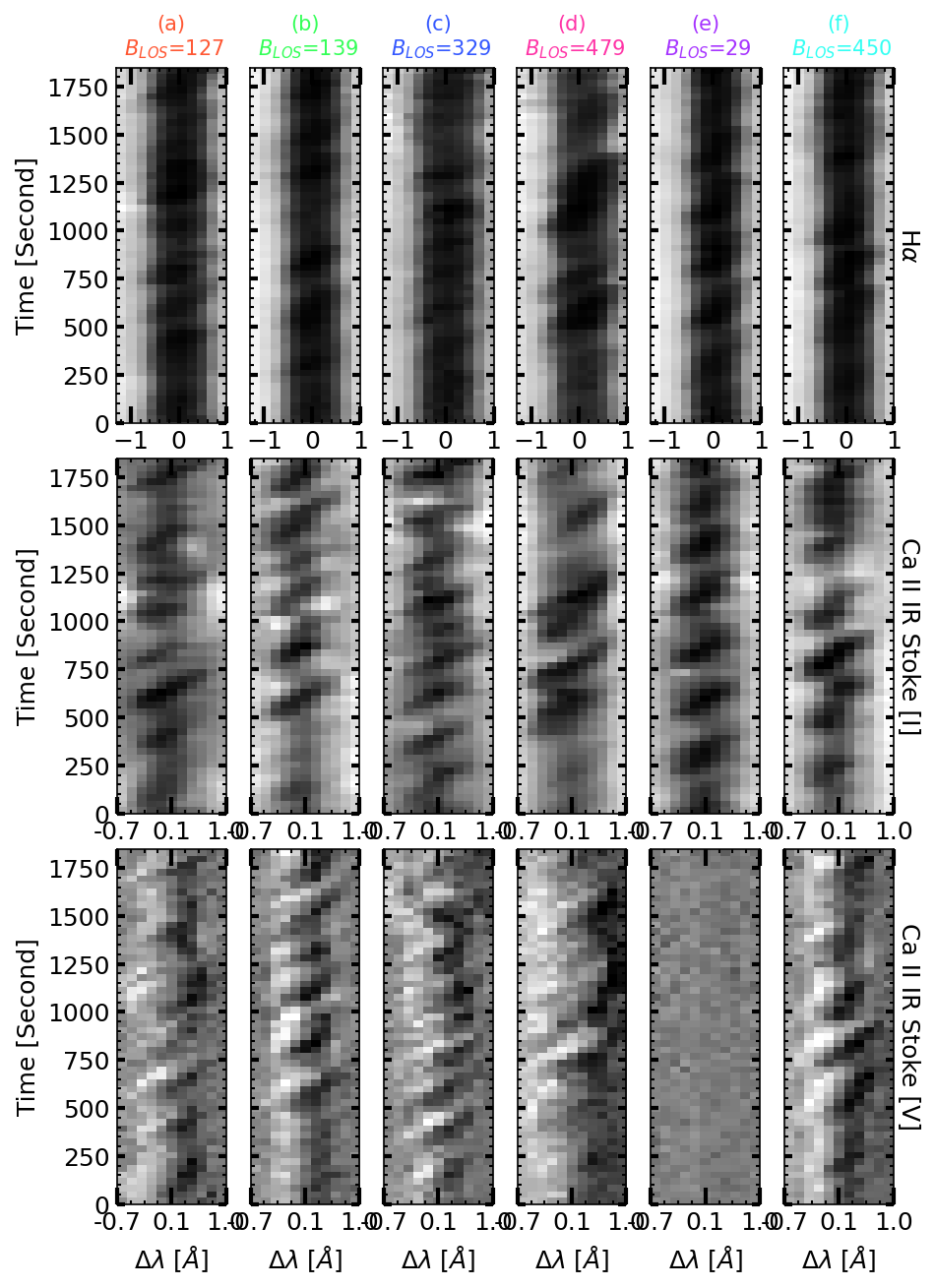}
	\caption{The right panel displays six columns of $\lambda$-t plots, each corresponding to a different location marked by colored X symbols in the left panel. Each column represents coordinated chromospheric observations, with the top row showing the Stokes I parameter of H$\alpha$, followed by the Stokes I and Stokes V parameters of the Ca II 8542 \AA~ spectral line. The left panel consists of two columns: the first shows the average chromospheric appearance in the blue wing of H$\alpha$ (H$\alpha$ - 0.8 Å), the average HMI magnetogram, and the coronal structure from the SDO/AIA 171 Å channel, while the second provides a zoomed-in view (cyan box) highlighting shock wave locations. The LOS magnetic field strength at each studied point is indicated atop the respective $\lambda$-t plot. Notably, when the photospheric field exceeds 100 G, shock waves are detected in both Stokes I and Stokes V parameters, whereas weaker fields (<30 G) lack a clear Stokes V signature, as seen in location (e) which was expected.}
	\label{Fig2}
\end{figure*}

The right panel of Figure \ref{Fig2} displays six columns of $\lambda$-t plots, each corresponding to a different location marked by distinct colored X symbols in the left panel of Figure \ref{Fig2}. Each column ($\lambda$-t) in the right panel represents coordinated observations of the chromosphere, where the top row shows the Stokes I parameter of H$\alpha$, followed by the Stokes I and Stokes V parameters of the Ca II 8542 \AA~ spectral line as one moves downward. The left panel of Figure \ref{Fig2} consists of two columns. The first column presents the average appearance of the chromosphere in the blue wing of H$\alpha$ (H$\alpha$ - 0.8 \AA), the average HMI magnetogram, and the average coronal structure observed in the SDO/AIA 171 Å channel, arranged from top to bottom for the entire observational period. The second column provides a zoomed-in view (highlighted by a cyan box) of the corresponding images in the first column, offering a clearer depiction of the shock wave locations across different layers of the solar atmosphere. In addition to these six points, five more locations have been selected that are co-spatial with the IRIS slit position. The results from these points are discussed separately in Section \ref{Sec3.2}.  

It is important to emphasize that although only six representative locations are presented here for clarity, many other points within the FoV also exhibit similar shock wave signatures. Additionally, these shocks appear comparatively weak in both chromospheric spectral lines when located in regions of strong magnetic field, such as within the dark pore. The distinct sawtooth patterns observed in these $\lambda$-t plots of both the chromospheric spectral lines suggest the presence of recurrent shock waves in the solar chromosphere. The LOS magnetic field strength of the photosphere at each studied point is indicated at the top of the respective $\lambda$-t plot.

\subsubsection{Possible association of shock wave locations with the average photospheric magnetic field and its inclination}\label{Sec3.1.1}

%\begin{figure*}
%	\centering
%	\includegraphics[width=180mm]{figures/Fig3a.png}
%	\caption{The left panel shows the average appearance of the chromosphere in the blue wing of Ca II 8542 \AA~ (Ca II 8542 \AA~ - 0.8 Å), with 50 selected shock wave locations marked by colored X symbols. The right panel presents histograms depicting the distribution of average magnetic field strength and inclination in the photosphere at these shock wave locations. These 50 shock waves are shown in Figure \ref{FigA1}. The results indicate that most shock waves are associated with photospheric magnetic field strengths between 0-400 G and inclinations around 50$^\circ$.}
%	\label{Fig3}
%\end{figure*}

\begin{figure*}
	\centering
	\includegraphics[width=180mm]{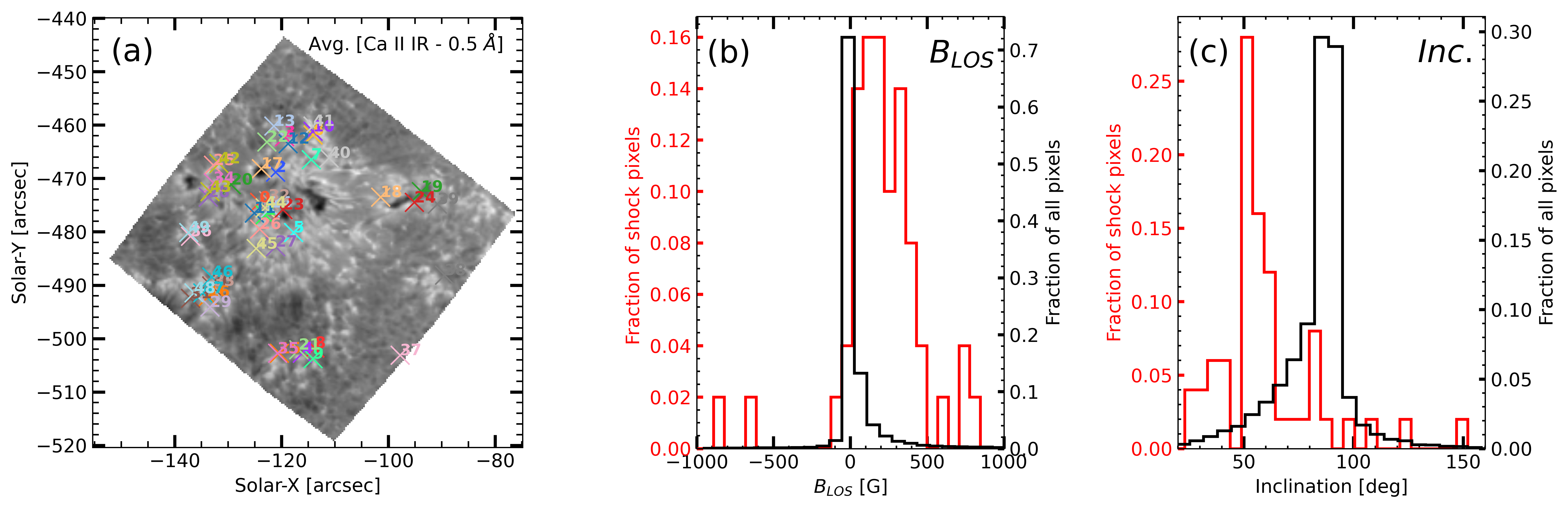}
	\caption{Panel (a) shows the average appearance of the chromosphere in the blue wing of Ca II 8542 \AA~ (Ca II 8542 \AA~ - 0.8 Å), with 50 selected shock wave locations marked by colored X symbols. Panels (b) and (c) show histograms of the average magnetic field strength and inclination in the photosphere, comparing the distributions at the identified shock wave locations with those across the entire FoV. These 50 shock waves are shown in Figure \ref{FigA1}.}
	\label{Fig3}
\end{figure*}

To investigate the potential association between shock wave locations in the chromosphere and the average photospheric magnetic field and its inclination, we identified 50 samples (marked by colored X symbol over the average appearance of the chormosphere in the blue wing of the Ca II 8542 \AA~ (Ca II 8542 \AA~ - 0.8 \AA) in  panel (a) of Figure \ref{Fig3}), where shock wave signatures were consistently observed throughout most of the observation period. It should be noted that among these 50 locations, the first 11 same locations as (a)-(k) marked over the left panel of Figure \ref{Fig2}. Panels (b) and (c) of Figure \ref{Fig3} present histograms illustrating the distribution of the average magnetic field strength and inclination in the photosphere at the shock wave locations, compared to the corresponding distributions for all pixels within the FoV. The shock wave signatures at these 50 locations in the Ca II 8542 \AA~ only are shown in Figure \ref{FigA1} as Ca II 8542 \AA~ present a more clearer saw tooth pattern than H$\alpha$ as the line core of the Ca II 8542 \AA~ is narrower as compared to the H$\alpha$.

\subsection{Shock waves in the transition region}\label{Sec3.2}

\begin{figure}
	\centering
	\includegraphics[width=81mm]{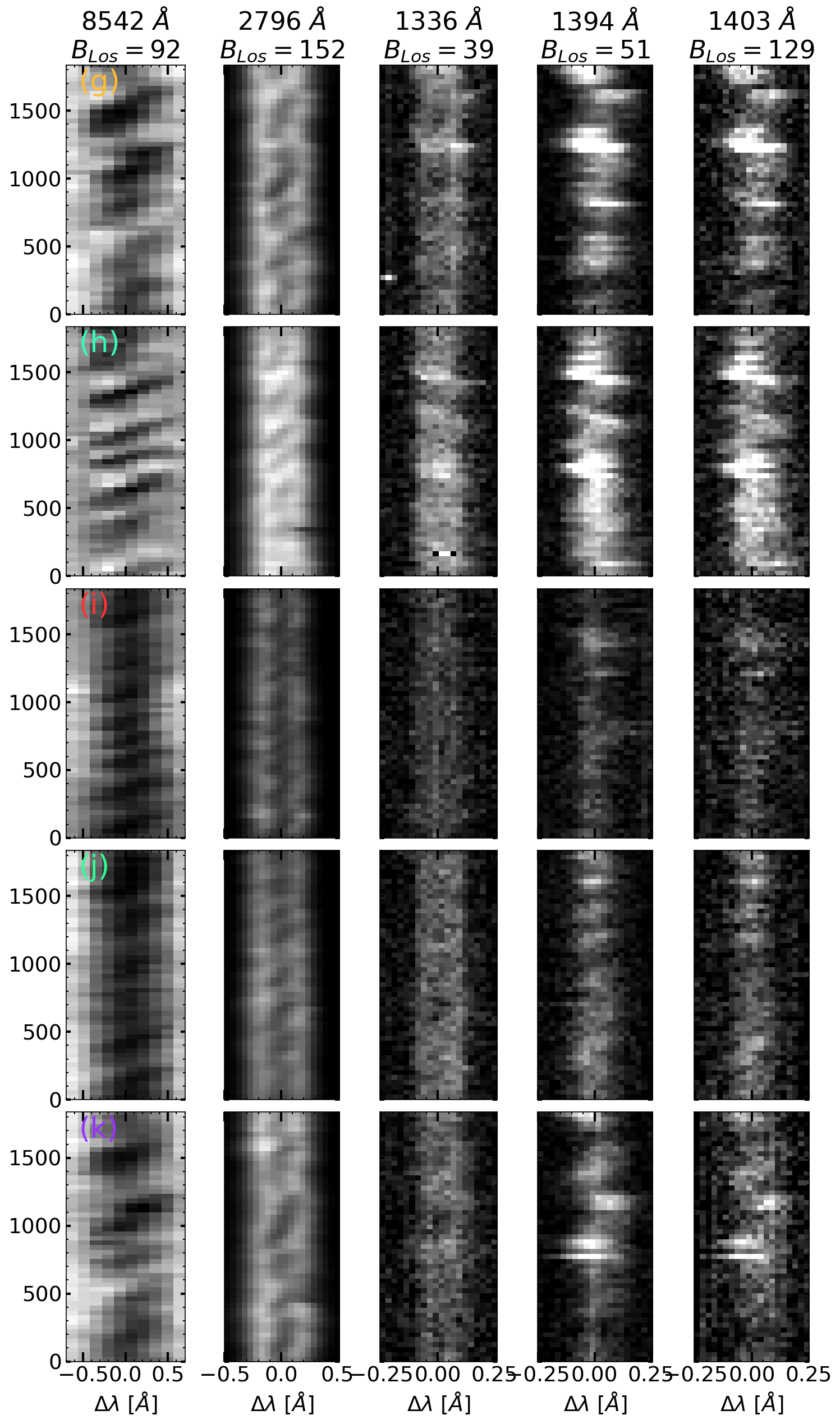}
	\caption{Five rows of $\lambda$-t plots for five selected locations ((g)--(k)), showing different atmospheric layers from left to right: the middle chromosphere (Ca II 8542 \AA), upper chromosphere (Mg II K 2796 Å), and transition region (C II 1336 Å, Si IV 1394 Å, and Si IV 1403 Å). The detection of shock wave signatures in transition region lines indicates their upward propagation through the solar atmosphere.}
	\label{Fig4}
\end{figure}

To investigate whether the shock waves observed in the chromosphere, reach transition region temperature regimes, we analyze coordinated spectra from both the chromosphere and transition region at locations overlapping the IRIS slit positions.

Figure \ref{Fig4} presents five rows of $\lambda$-t plots for five selected locations ((g)--(k)), capturing different atmospheric layers from left to right: the middle chromosphere (Ca II 8542 \AA), upper chromosphere (Mg II K 2796 Å), and transition region (C II 1336 Å, Si IV 1394 Å, and Si IV 1403 Å). The presence of shock wave signatures in transition region lines suggests their upward propagation through the solar atmosphere. It is important to note that while many other locations along the slit exhibited similar behavior, only five representative cases are shown here for brevity.

\begin{figure}
	\centering
	\includegraphics[width=85mm]{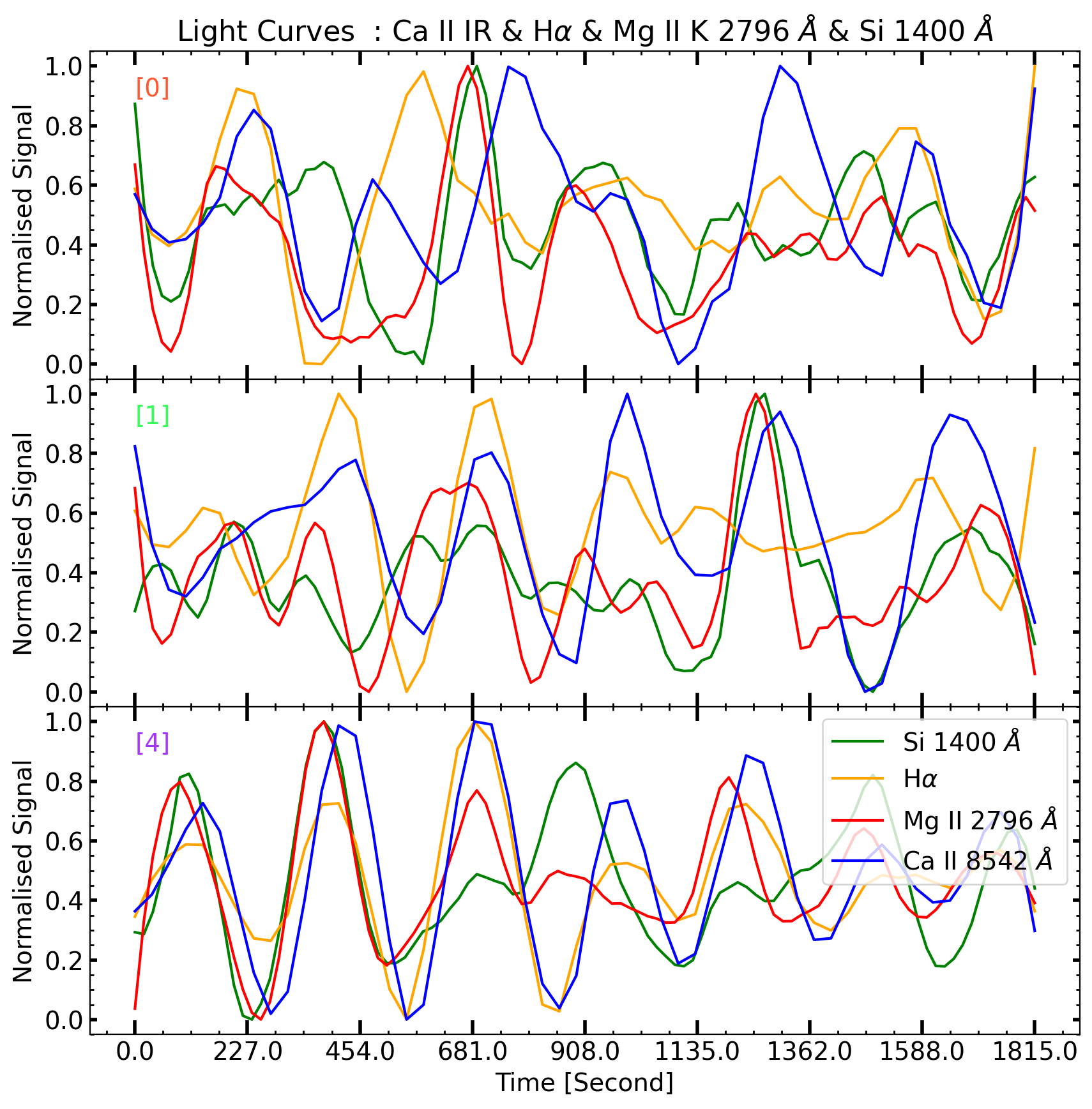}
	\caption{Light curves from the mid chromosphere (H$\alpha$ and Ca II 8542 \AA~ line centers), upper chromosphere (Mg II K 2796 \AA) and the transition region (Si IV 1400 Å) for three selected locations. These locations, marked by X and labeled as (a), (b), and (e) in the left panel of Figure \ref{Fig2}, correspond to locations [0], [1], and [4] in panel (a) of Figure \ref{Fig3}.}
	\label{Fig5}
\end{figure}

The above examples suggest that shock waves generated in the chromosphere propagate upward, at least to the transition region. This indicates that, at the locations where shock waves are present, the transition region is coupled with the middle chromosphere. Consequently, one would expect the light curves from the chromosphere and the transition region at these locations to exhibit a correlation. Figure \ref{Fig5} presents light curves from the chromosphere (line center of Ca II 8542 \AA~ and H$\alpha$), upper chromopshere (Mg II 2796 \AA) and the transition region (Si IV 1400 Å) for three selected locations marked by X and labeled as (a), (b), and (e) in the left panel of Figure \ref{Fig2} that corresponds to the same locations as [0], [1], and [4] panel (a) of Figure \ref{Fig3}.

\subsection{Shock waves in the Corona}\label{Sec3.3}

So far, we have investigated the propagation of chromospheric shock waves into the transition region. However, we are also interested in determining whether these shock waves extend beyond the transition region into coronal temperature regimes. Since we lack spectral observations corresponding to coronal temperatures and instead have coronal images, we analyze the correlation between light curves from the chromosphere, transition region, and corona to track the possible upward propagation of these shock waves.

\begin{figure}
	\centering
	\includegraphics[width=85mm]{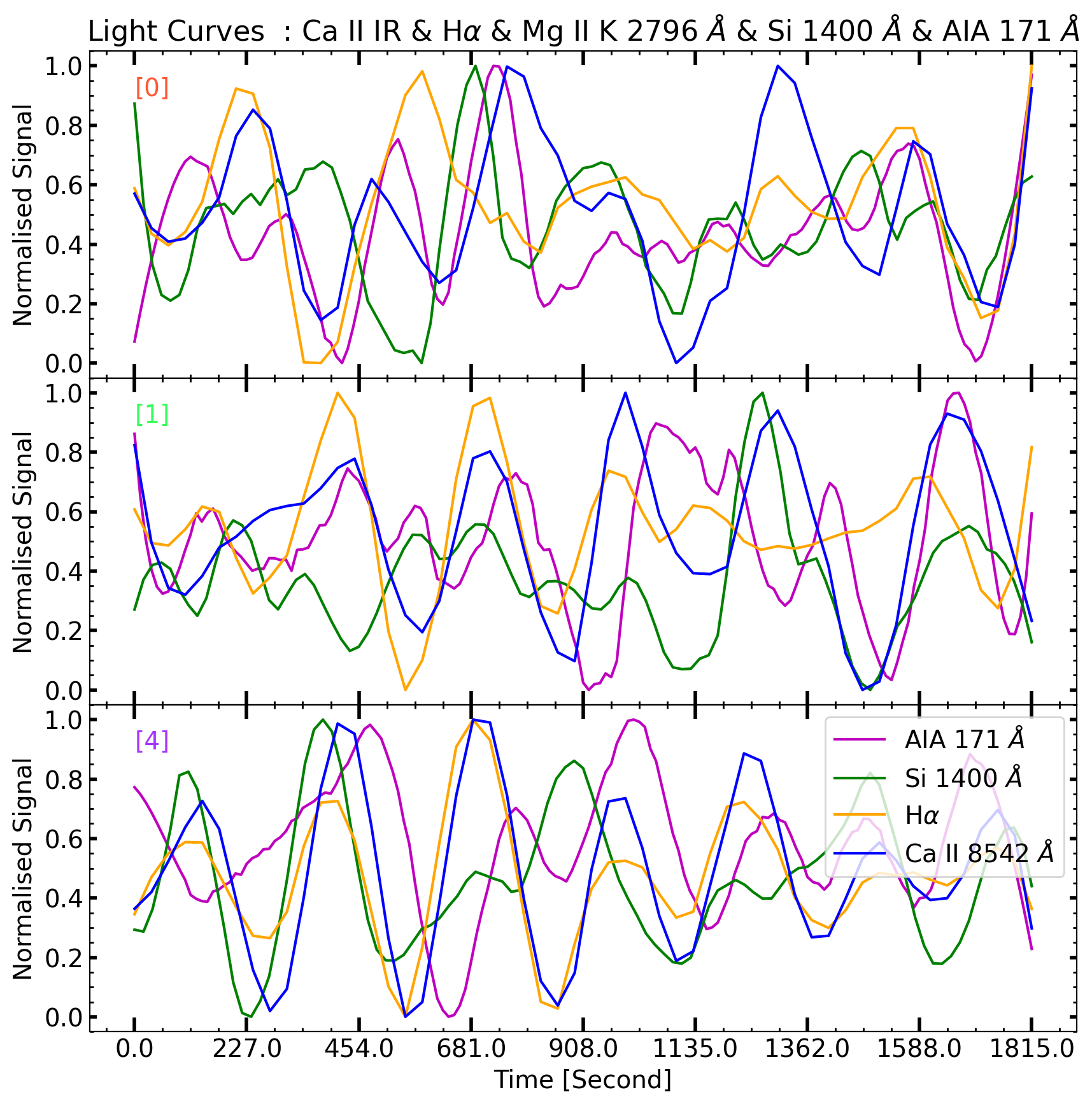}
	\caption{Light curves from the mid chromosphere (H$\alpha$ and Ca II 8542 \AA~ line centers), transition region (Si IV 1400 Å) and the corona (AIA 171 \AA) for the three points shown in Figure \ref{Fig5}.}
	\label{Fig6}
\end{figure}

\begin{figure}
	\centering
	\includegraphics[width=85mm]{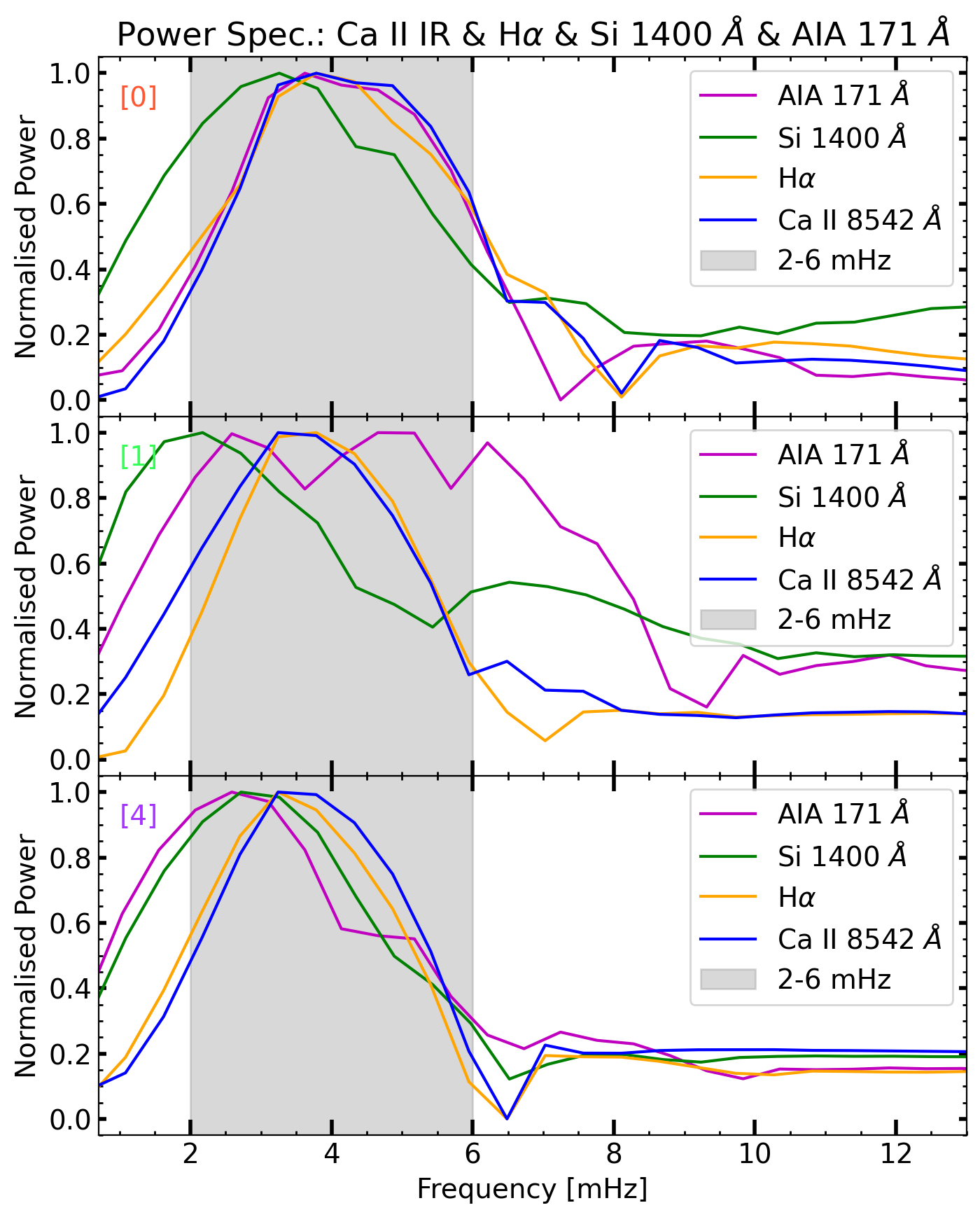}
	\caption{The power spectrum of the light curves of corresponding to the chromosphere (blue and orange),transition region (green) and the corona (magenta).}
	\label{Fig7}
\end{figure}

Figure \ref{Fig6} displays the light curves from the chromosphere, transition region, and corona for the same locations as shown in Figure \ref{Fig5}. Although we examined light curves from both AIA 193 Å~ and AIA 171 Å, we present only the AIA 171 Å light curve to represent the corona, as both channels exhibited nearly identical variation patterns. To investigate the wave properties at different atmospheric heights, we applied a Fourier transform to these light curves and estimated the power spectra, which are presented in Figure \ref{Fig7}. A similar analysis was performed for 15 additional locations, marked by X in different colors in panel (a) of Figure \ref{Fig3}. The results for these locations are shown in Figure \ref{FigA2}, to reinforce the scenario that at least some of the nearly 3.5 mHz oscillations observed in AIA intensity images are manifestations of chromospheric shock waves, as discussed in Section \ref{Sec4}.

\subsection{Shock wave flux determination and comparison with the radiative lossess of the chromosphere}\label{Sec3.4}

Our analysis supports the scenario that shock waves originate in the chromosphere and propagate upward through the solar atmosphere. To understand their role in energy transport, we estimated their shock wave flux and examined whether they contribute to energy deposition in the chromosphere. The energy deposition due to shock waves can be determined by calculating the difference in shock wave power flux at two different heights in the chromosphere.  

We estimate the shock wave power flux at a given height using the formulation from previous studies \citep{2013A&A...560A..84S,2016ApJ...826...49S,2020ApJ...890...22A}:  

\begin{equation}
	F_{\rm ac,tot} = \int_{\nu_{\rm ac}}^{\nu_{\rm max}} \rho P_{\rm v}(\nu) \frac{\upsilon_{\rm gr}(\nu)}{TF(\nu)} \, d\nu,  
	\tag{1}
\end{equation}

where:  

\begin{itemize}

\item \(\rho\) is the gas density,  
\item  \(P_{\rm v}(\nu)\) is the spectral power density derived from the velocity signal,  
\item  \(\upsilon_{\rm gr}(\nu)\) is the group velocity of wave energy transport, given by:

\end{itemize}

\begin{equation}
	\upsilon_{\rm gr} = c_s \sqrt{1 - \left(\frac{\nu_{\rm ac}}{\nu}\right)^2}.  
	\tag{2}
\end{equation}

Here, \(c_s\) is the sound speed, defined as:

\begin{equation}
	c_s = \sqrt{\frac{\gamma p}{\rho}},  
	\tag{3}
\end{equation}

where \(\gamma\) is the adiabatic index ($ \gamma $=5/3 for mono-atomic gas) and \(p\) is the gas pressure.  

The acoustic cutoff frequency, \(\nu_{\rm ac}\), which determines the lowest frequency at which waves can propagate, is given by:

\begin{equation}
	\nu_{\rm ac} = \frac{\gamma g \cos\theta}{4\pi c_s},  
	\tag{4}
\end{equation}

where \(g\) is the gravitational acceleration ($ g $=274 m s$^{-2}$ at the surface of the Sun) and \(\theta\) is the inclination of the magnetic field at the photosphere.  

In Eq. (1), \(TF(\nu)\) represents the transfer function of the solar atmosphere \citep{2009A&A...508..941B}. If \(TF(\nu) = 1\), it means that the observed Doppler signal captures all waves at a given frequency without loss. However, when \(TF(\nu) < 1\), some signal is lost due to the finite height range of spectral-line contribution functions, especially affecting high-frequency waves. While accurately determining \(TF(\nu)\) requires a detailed atmospheric model, in this study, we assume \(TF(\nu) = 1\) for simplicity. If the actual \(TF(\nu)\) were considered, the estimated total shock wave energy flux could be higher.

To obtain the physical parameters of the solar chromosphere that include velocity, gas pressure, and density and other physical parameters, we employ the STiC inversion code. STiC is an MPI-parallel inversion tool used to retrieve the vertical stratification of atmospheric parameters thus helps in monitoring the evolution of specific chromospheric features. STiC is based on a modified version of the RH radiative transfer code \citep{2001ApJ...557..389U} and uses cubic Bezier solvers to solve the polarized radiative transfer equation \citep{2013ApJ...764...33D}. Operating under non-LTE conditions with the assumption of statistical equilibrium, STiC is capable of fitting multiple spectral lines simultaneously. It includes an efficient approximation to account for partial redistribution effects \citep{2012A&A...543A.109L}. The code assumes a plane-parallel geometry for each pixel (1.5D approximation) and incorporates an LTE equation of state from the Spectroscopy Made Easy (SME) library \cite{2017A&A...597A..16P}. STiC performs the inversion by iteratively adjusting atmospheric parameters such as temperature ($T$), line-of-sight velocity ($V_{\rm LOS}$), and microturbulence ($V_{\rm turb}$), in order to minimize the chi-squared ($\chi^2$) difference between the synthetic and observed spectral profiles. The inferred atmospheric parameters are stratified with respect to the logarithmic optical depth at 500~nm, denoted as $\log \tau_{500}$. Gas pressure and density stratifications are computed under the assumption of hydrostatic equilibrium.

Perturbations to the model atmosphere are introduced at specific locations along the $\log \tau_{500}$ scale, referred to as nodes, and are interpolated over the full depth grid. In our analysis, we inverted the full Stokes parameters of the Ca~\textsc{ii}~8542~\AA\ spectral line, using the Fontenla, Avrett, and Loeser (FAL-C; \cite{1993ApJ...406..319F}) model atmosphere as the initial guess. A six-level Ca~\textsc{ii} atom model with nine nodes for temperature and velocity and five nodes for the magnetic field components have been used for the inversion. The average response height of the line core of Ca~\textsc{ii} 8542 \AA~  was found to be at $\log \tau_{500} =-4.8 $.

While STiC does not directly provide uncertainties in the inferred physical parameters —such as temperature, $V_{\rm LOS}$, $V_{\rm turb}$, and the magnetic field components— these uncertainties can be estimated indirectly using the formalism outlined by \citet{2024ApJS..271...24S}, expressed as:

\begin{equation}
		\sigma_{p}^{2} = \frac{2}{nm + r} \cdot 
		\frac{
			\sum\limits_{i=1}^{q} \left[ I^{\text{obs}}(\lambda_{i}) - I^{\text{syn}}(\lambda_{i}; \mathbf{M}) \right]^{2} \cdot \frac{w_{i}^{2}}{\sigma_{i}^{2}}
		}{
			\sum\limits_{i=1}^{q} R_{p}^{2}(\lambda_{i}) \cdot \frac{w_{i}^{2}}{\sigma_{i}^{2}}
		}
		\tag{5}
\end{equation}

\noindent where:

\begin{itemize}
		\item \( \sigma_p^2 \): Variance of the model parameter \( p \).
		\item \( nm + r \): represents the total number of free parameters in the inversion. Here, \( n \) is the number of nodes along the optical depth for each of the \( m \) atmospheric variables, and \( r \) denotes the number of parameters that remain constant with depth (e.g., the upper boundary gas pressure, where \( r = 1 \)).
		\item \( q \): Total number of wavelength points.
		\item \( I^{\text{obs}}(\lambda_i) \): Observed intensity at wavelength \( \lambda_i \).
		\item \( I^{\text{syn}}(\lambda_i; \mathbf{M}) \): Synthetic intensity at wavelength \( \lambda_i \) from model
		atmosphere \( \mathbf{M} \).
		\item \( \mathbf{M} \): Vector of model parameters (e.g., temperature, velocity, magnetic field).
		\item \( w_i \): Weight assigned to wavelength \( \lambda_i \).
		\item \( \sigma_i \): Uncertainty (standard deviation) in the observed intensity at \( \lambda_i \).
		\item \( R_p(\lambda_i) = \frac{\partial I^{\text{syn}}(\lambda_i; \mathbf{M})}{\partial p} \): represents the response function of the intensity to a
		perturbation in the physical parameter \textit{p} of the model
		atmosphere \( \mathbf{M} \).
\end{itemize}

An example of the velocity stratification for both quiet sun regions and the time average of the shock wave locations (labeled as (a) in Figure \ref{Fig2}), along with the associated uncertainties estimated using the above formulation, is presented in Figure~\ref{FigA3}.

Utilizing the Doppler velocity, gas pressure, and mass density from the STiC inversion, in conjunction with the photospheric inclination obtained from SDO/HMI, we estimate the shock wave energy flux at two chromospheric layers, corresponding to $\log\tau_{5000} = -5.5$ and $\log\tau_{5000} = -4.5$. These layers are chosen based on the sensitivity of the inverted Doppler velocities to these heights, as supported by \citet{2016MNRAS.459.3363Q}. The energy flux calculation is performed over 50 selected locations where shock waves are identified in the chromosphere (see Figure~\ref{Fig3}).

\begin{figure}
	\centering
	\includegraphics[width=81mm]{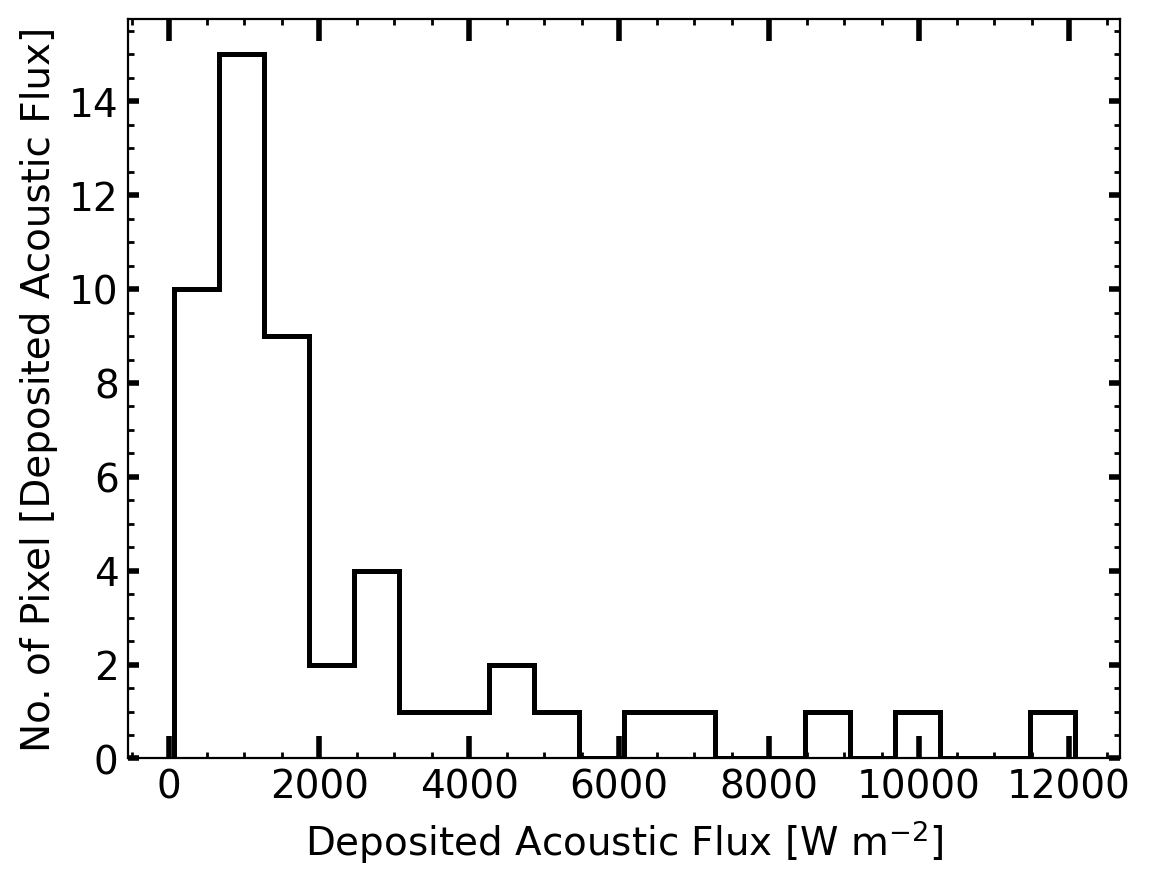}
	\caption{Histogram displaying the total deposited flux, calculated as the difference in shock wave power flux between $\log\tau_{5000} = -5.5$ and $\log\tau_{5000} = -4.5$, across 50 chromospheric locations where shock waves were identified.}
	\label{Fig8}
\end{figure}

A histogram depicting the total deposited flux (defined as the difference in energy flux between the two heights) for the corresponding shock wave pixels is shown in Figure~\ref{Fig8}. It is to be noted that we could not compute the uncertainties in the deposited flux since errors in gas pressure and mass density could not be derived, as they are not free parameters in the STiC inversions.

\section{Summary and Conclusion}  \label{Sec4}
	
Shock waves are frequently observed in bright regions and sunspot umbrae, spanning chromospheric to transition region temperatures~(\cite{1997ApJ...481..500C,2003A&A...403..277R,2015ApJ...803...44M,2021ApJ...906..121K}). Although many studies have focused on individual atmospheric layers, coordinated multi-height observations are limited, and it remains uncertain whether shocks from the lower chromosphere reach the transition region. Understanding their propagation is vital, as they may transport energy upward and contribute to chromospheric and coronal heating. In this study, we use coordinated SST, IRIS, and SDO observations to track shock wave evolution across multiple heights in the solar atmosphere.

To study shock waves in the chromosphere, we analyze the Stokes I parameter of the H$\alpha$ line and both Stokes I and Stokes V parameters of the Ca II 8542 \AA~ spectral line. Our results suggests that shock waves appear as sawtooth patterns in $\lambda$-t plots of both Stokes I and Stokes V parameters when the average magnetic field in the photosphere is sufficiently strong (Figure \ref{Fig2}). Notably, when the photospheric magnetic field exceeds 100 G, shock waves are detectable in both Stokes I and Stokes V parameters, with Stokes V representing circular polarization. However, at locations where the photospheric field is relatively weak (<30 G, typical of quiet Sun (QS) regions), shock signatures are absent in the Stokes V parameter of the Ca II 8542 \AA~ line. This suggests that shock waves can also be associated with QS regions in the solar atmosphere. Interestingly, some shock wave locations (e.g., regions (a) and (f) of Figure \ref{Fig2}) lie in close proximity to coronal loop footpoints. If these waves propagate upward, they could have significant implications for the dynamics and heating of coronal loops.

In order to study the possible association of these shock waves with the photospheric magnetic field and inclination we studied the 50 clean shock wave location and their histogram reveal that most of the shock waves are associated with average magnetic field strengths between 0-400 G and magnetic field inclinations around 50$^\circ$. It is important to note that we excluded locations where the shock wave signatures were diffuse or only present for a brief portion of the observation period. Therefore including these locations could alter the histogram results. On the other hand, the comparison of these shock wave signature (saw-tooth patterns in the $\lambda$-t plots) associated with different locations in the chromosphere reveals that shock waves can exhibit varying amplitudes, as indicated by the maximum excursion on the blue wing of the chromospheric spectral lines. This variation in amplitude may suggests differences in shock wave energy, propagation characteristics, and interaction with the local magnetic field and atmospheric conditions.

The traces of the signatures of these shock waves were also observed in coordinated spectral data from the chromosphere and transition region temperature regimes, though not with the same one-to-one correspondence of individual shock waves as observed in Mg~II~K~2796~\AA\ and Ca~II~8542~\AA. However, in some cases, even though the signature of the shock wave is clearly visible in the Ca II 8542 \AA~ and Mg II K 2796 \AA~ lines, there is no clear signature but rather blurred in the transition region tempearure regimes. This behavior is still consistent with the expected shock wave propagation: as the wave encounters the transition region, the rise in temperature leads to an increase in the local sound speed, causing partial reflection. In addition, thermal conduction—negligible in the chromosphere but significant in the transition region—may further dissipate the shock energy, thereby lowering its amplitude and making its signature less pronounced. However, a definitive conclusion requires clearer and more consistent shock wave signatures in such coordinated observations, as their upward propagation may often be hindered by the presence of more horizontal magnetic canopies (\cite{2009A&A...494..269V}).

For locations outside the IRIS slit coverage, we correlated chromospheric and transition region light curves. Despite different cadences, these light curves exhibit good correlations, further supporting the presence of propagating shock waves. To assess the extent of shock wave propagation, we correlated light curves from the corona with those from the chromosphere and transition region (Figure \ref{Fig5} and Figure \ref{Fig6}). Additionally, we computed the power spectra of these light curves to identify waves with dominant frequency (Figure \ref{Fig7} and Figure \ref{FigA2}). Power spectra of all the light curves show a dominat power around 3 mHz for the most of the identified pixels. However, the observed out-of-phase behavior between the light curves, along with deviations in the correlation between the light curves and the peak power, may arise either from differences in the line-formation characteristics of the spectral lines used or from abrupt changes in the atmospheric parameters caused by shock wave propagation.

However, to confirm whether these oscillations originate from the lower chromosphere, long-duration observations with high-frequency resolution in the chromosphere are required, forming a key direction for future work. High-resolution observations from  the 4 m Daniel K. Inouye Solar Telescope (DKIST, \cite{2020SoPh..295..172R}) in chromospheric passbands, as well as 
future missions like the Multi-slit Solar Explorer (MUSE, \cite{2022ApJ...926...52D}), which will provide high-resolution spectral data of coronal passbands, could provide valuable insights into the propagation of shock waves into coronal temperatures and their possible contribution to heating the solar atmosphere.

In conclusion, our study demonstrates that shock waves propagate upward through the solar atmosphere, as evidenced by their characteristic sawtooth wave pattern in the $\lambda$-t plots of chromospheric spectral lines and in the spectral lines corresponding to the transition region temperature regime. Additionally, our findings suggest that these shock waves may extend beyond the transition region, as the light curves from the chromosphere, transition region, and coronal passbands exhibit good correlations. Since these shock waves originate in the low chromosphere and propagate upward, they are magneto-acoustic in nature, as the rapid decrease in gas pressure with height causes the magnetic field to expand, effectively filling all the space about a few hundred kilometers above the photosphere (\cite{2019ARA&A..57..189C}). 

Our analysis also indicates that these shock waves are closely linked to bright regions in the chromosphere, which are also associated with the origin of spicules. This finding paves the way for our future work (Chaurasiya et al., in prep.), where we observationally investigate the relationship between the origin of spicules with the presence of shock waves and the associated propagating coronal disturbances. Understanding this connection could provide deeper insights into the dynamic coupling between the lower and upper solar atmosphere and the role of shock waves in energy transport.

On the other hand, our estimates of shock wave flux deposition suggest that shock waves may deposit a significant amount of energy flux into the chromosphere highlights their potential role in maintaining the energy balance of the chromosphere. However, the energy from shock waves alone doesn’t seem to be enough, indicating other processes must also be at work, continuously supplying energy to make up for the losses. Identifying these supplementary energy sources are crucial for a comprehensive understanding of chromospheric heating processes.

\section*{Acknowledgements}

We thank the referee for his/her comments and suggestions. The Swedish 1-m Solar Telescope is operated on the island of La Palma by the Institute for Solar Physics of Stockholm University in the Spanish Observatorio del Roque de los Muchachos of the Instituto de Astrofísica de Canarias. The Swedish 1-m Solar Telescope, SST, is co-funded by the Swedish Research Council as a national research infrastructure (registration number 4.3-2021-00169). We would like to thank SST team for making the data publicly available. RE acknowlages the NKFIH OTKA (Hungary, grant No. K142987). RE is also grateful to Science and Technology Facilities Council (STFC, grant No. ST/M000826/1) UK, acknowledges  PIFI (China, grant number No. 2024PVA0043) and the NKFIH (Hungary)  Excellence Grant (grant nr TKP2021-NKTA-64) for enabling this research. This work was also supported by the International Space Science Institute project (ISSI-BJ ID 24-604) on ”Small-scale eruptions in the Sun”. We also thank Dr. Rahul Yadav for providing insights in utilising the STiC. We have made much use of NASA’s Astrophysics Data System Bibliographic Services.

%%%%%%%%%%%%%%%%%%%%%%%%%%%%%%%%%%%%%%%%%%%%%%%%%%
\section*{Data Availability}

The observational data used in this study was obtained from the Swedish Solar Telescope (SST) and is available online at the following link: \url{https://dubshen.astro.su.se/sst_archive/search}. \newline

\bibliographystyle{mnras}
\bibliography{shock_propagationv2} % if your bibtex file is called example.bib

\appendix 	
\label{appex}

\section{50 shock waves in the chromosphere}

This section presents the result of the shock waves which are observed at 50 different locations in the chromosphere marked with different color (labeled by different number) in panel (a) of Figure \ref{Fig3}. The histogram presenting the relation of these shock waves with photospheric magnetic field and inclination is shown in panel (b) and (c) of Figure \ref{Fig3} respectively.

\begin{figure}
	\centering
	\includegraphics[width=81mm]{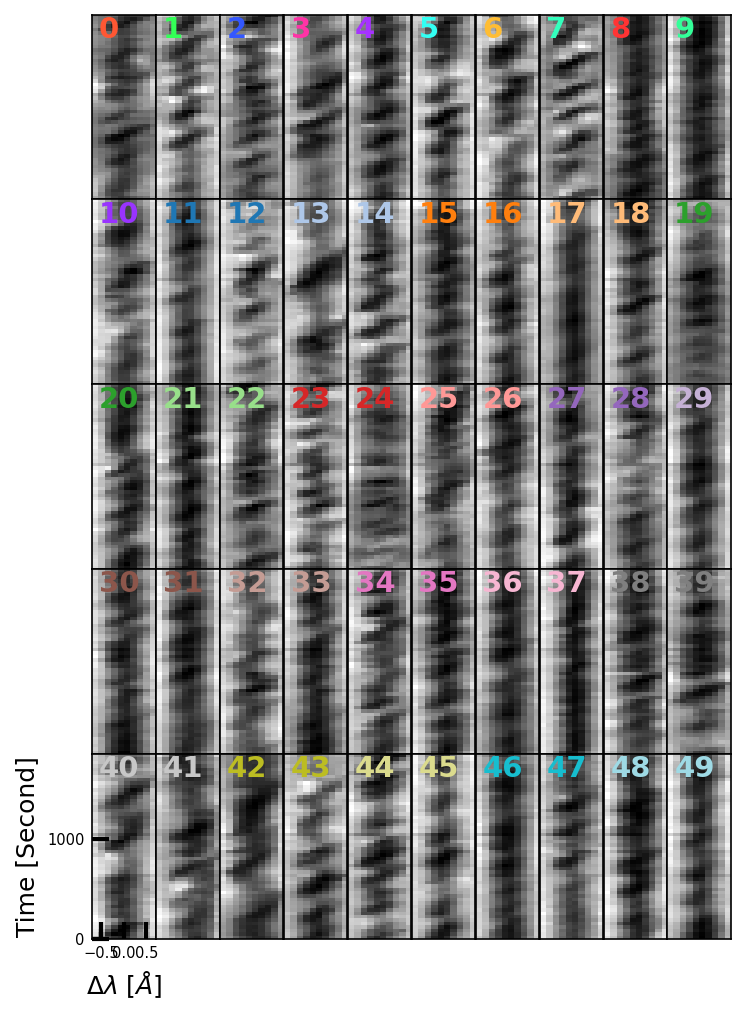}
	\caption{The above panel depicts the $\lambda$-t plots of the 50 different locations in the chromosphere marked by "X" in panel (a) of Figure \ref{Fig3}.}
	\label{FigA1}
\end{figure}

\section{Power Spectrum of the light curves corresponding to different heights in the solar atmosphere}

Similar to the 3 power spectrums shown in Figure \ref{Fig7}, this section presents 15 more such power spectrum corresponding to different locations marked different color (labeled by different number) in panel (a) of Figure \ref{Fig3}.

\begin{figure*}
	\centering
	\includegraphics[width=180mm]{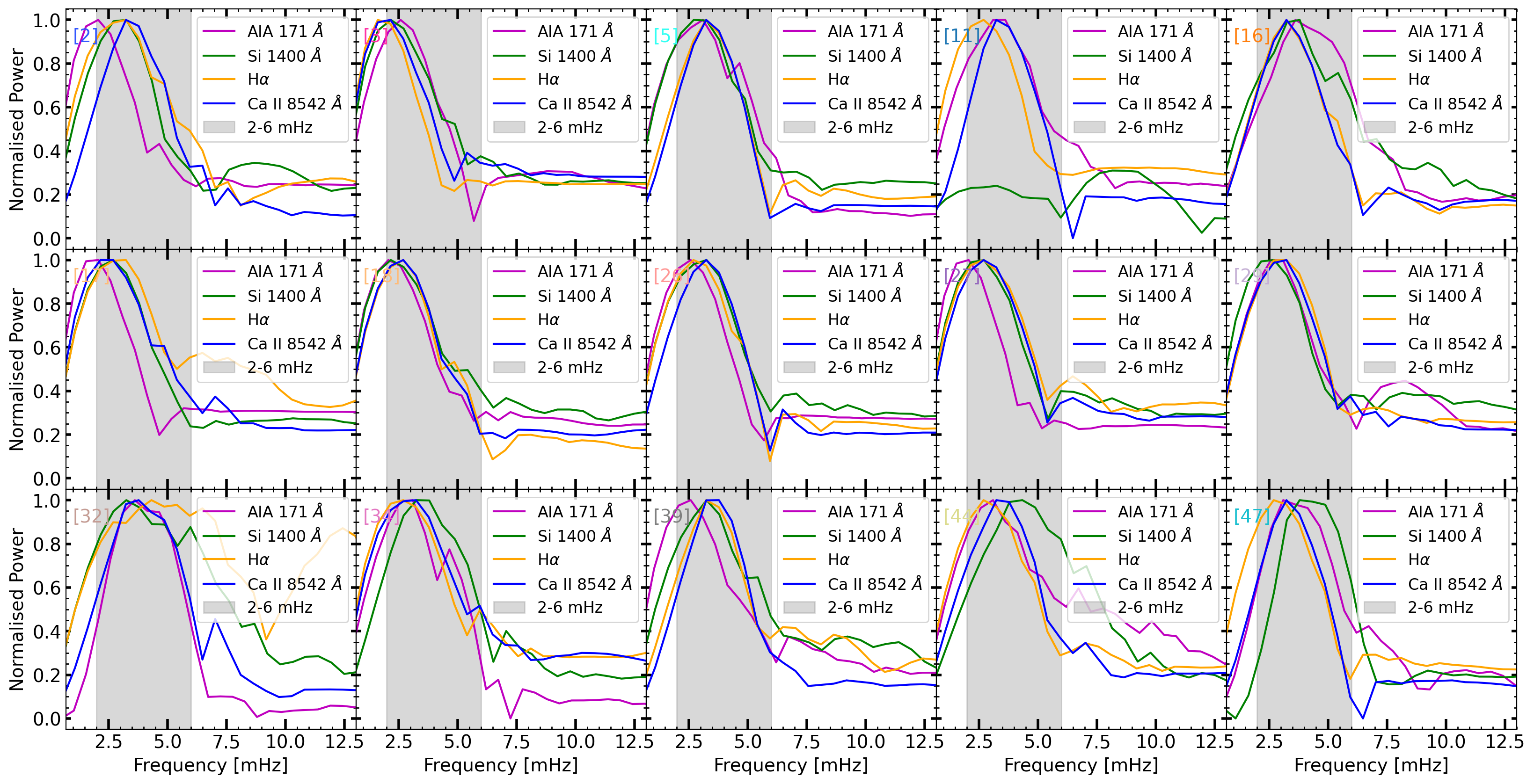}
	\caption{The power spectrum of the light curves corresponding to the chromosphere (Ca II 8542 \AA~ and H$\alpha$ line centers),transition region (Si IV 1400 \AA) and the corona (AIA 171 \AA) for the 15 points shown by X in panel (a) of Figure \ref{Fig3}.}
	\label{FigA2}
\end{figure*}

\section{Velocity Stratification of the Quiet-Sun Region and a Time-Averaged Shock Wave Location}
This section presents the V$_{LOS}$ stratification derived from STiC inversion for both a representative quiet-Sun region and a time-averaged shock wave location (labeled as (a) in Figure~\ref{Fig2}). The vertical green lines indicate the uncertainties associated with the inferred velocities. The two red dashed lines mark the atmospheric heights corresponding to $\log\tau_{5000} = -5.5$ and $\log\tau_{5000} = -4.5$, where the shock wave energy flux is evaluated.

\begin{figure}
    \centering
    \includegraphics[width=90mm]{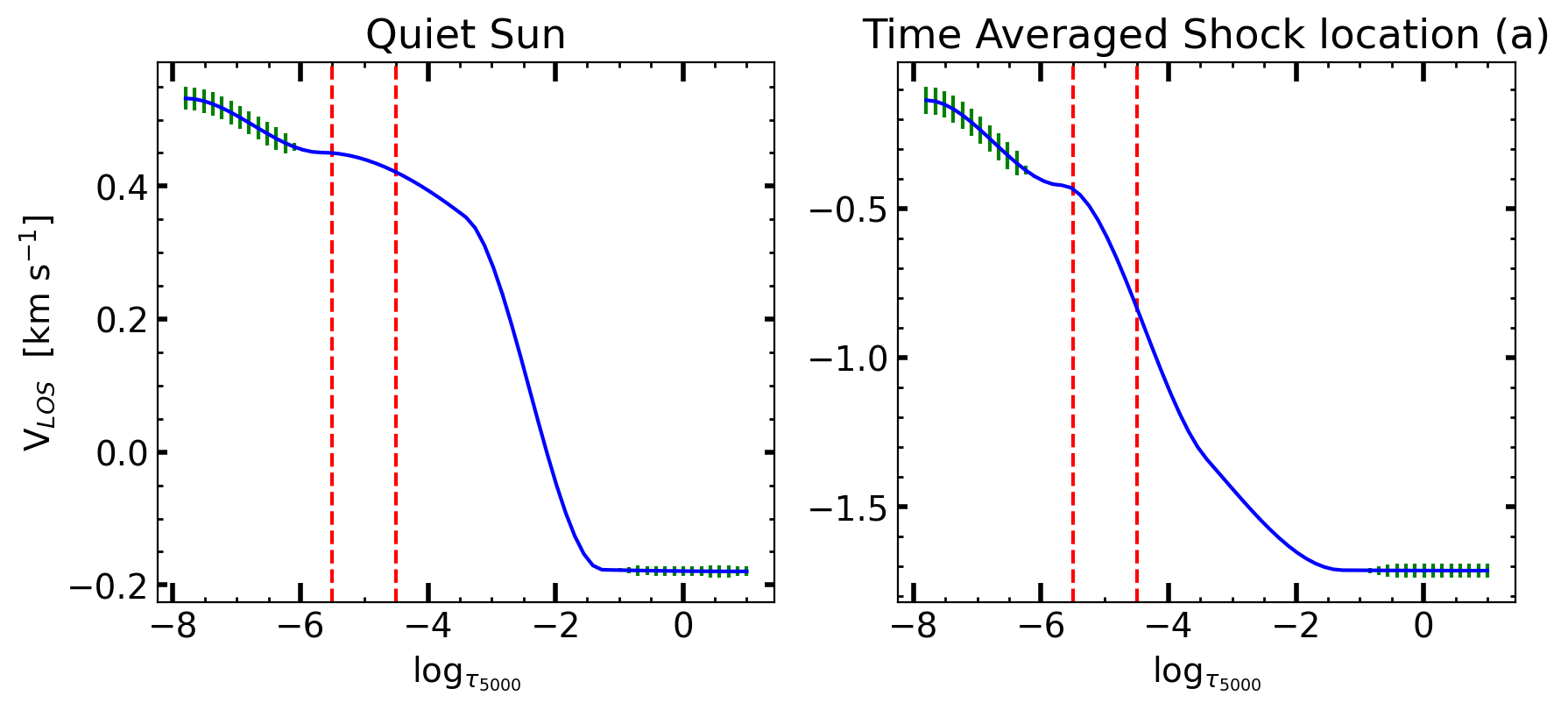}
    \caption{V$_{LOS}$ stratification for a quiet-Sun region and a time-averaged shock wave location (labeled as (a) in Figure~\ref{Fig2}). The vertical green lines represent uncertainties in the inferred velocities from the \texttt{STiC} inversion. The two red dashed lines indicate the atmospheric heights ($\log\tau_{5000} = -5.5$ and $-4.5$) at which the shock wave power flux has been estimated.}
    \label{FigA3}
\end{figure}

\bsp	% typesetting comment
\label{lastpage}
\end{document}